\documentclass[sigconf]{acmart}

\usepackage{booktabs} 

\usepackage{graphicx}
\usepackage{subfig}
\usepackage{url}
\usepackage{multirow}

\usepackage{amsmath,bm,amssymb}
\usepackage{algorithm}
\usepackage{algpseudocode}
\usepackage{varwidth}

\usepackage[flushleft]{threeparttable}
\algdef{SE}[DOWHILE]{Do}{doWhile}{\algorithmicdo}[1]{\algorithmicwhile\ #1}%

\usepackage{algorithm} 

\usepackage{mathtools}
\usepackage{url}
\usepackage{amsmath}
\newsavebox{\measurebox}
\renewcommand{\footnotesize}{\small}
\usepackage{amsfonts}
\usepackage[flushleft]{threeparttable}
\usepackage{url}

\usepackage{amssymb}

\usepackage{tikz}
\usetikzlibrary{calc}

\usepackage{colortbl}
\usepackage{booktabs}
\newcolumntype{L}[1]{>{\raggedright\let\newline\\\arraybackslash\hspace{0pt}}m{#1}}
\newcolumntype{C}[1]{>{\centering\let\newline\\\arraybackslash\hspace{0pt}}m{#1}}

\newcommand{\revised}[1]{\textcolor{black}{#1}}

\copyrightyear{2019} 
\acmYear{2019} 
\setcopyright{acmcopyright}
\acmConference[WSDM '19]{The Twelfth ACM International Conference on Web Search and Data Mining}{February 11--15, 2019}{Melbourne, VIC, Australia}
\acmBooktitle{The Twelfth ACM International Conference on Web Search and Data Mining (WSDM '19), February 11--15, 2019, Melbourne, VIC, Australia}
\acmPrice{15.00}
\acmDOI{10.1145/3289600.3290975}
\acmISBN{978-1-4503-5940-5/19/02}
\begin{document}
\title{A Simple  Convolutional  Generative Network for Next Item Recommendation}

\author{Fajie Yuan}
\authornote{\scriptsize A Part of work was done while at Telefonica Research, Spain and University of Glasgow, UK.}
\affiliation{%
	\institution{Tencent}
	\city{Shenzhen} 
	\state{China} 
}
\email{fajieyuan@tencent.com }

\author{Alexandros Karatzoglou} 
\affiliation{%
	\institution{Telefonica Research}
	\city{Barcelona} 
	\state{Spain} 
}
\email{alexandros.karatzoglou@gmail.com}

\author{Ioannis Arapakis} 
\affiliation{%
	\institution{Telefonica Research}
	\city{Barcelona} 
	\state{Spain} 
}
\email{arapakis.ioannis@gmail.com}

\author{Joemon M Jose} 
\affiliation{%
	\institution{University of Glagow	}
	\city{Glasgow} 
	\state{UK} 
}
\email{joemon.jose@glasgow.ac.uk}

\author{Xiangnan He}
\affiliation{%
	\institution{National University of Singapore}
	\country{Singapore}
}
\email{xiangnanhe@gmail.com}

\renewcommand{\shortauthors}{F. Yuan et al.}

\begin{abstract}
	Convolutional Neural Networks (CNNs) have been recently introduced in the domain of session-based next item recommendation. An ordered collection of past items the user has interacted with in a session (or sequence) are embedded into a 2-dimensional latent matrix, and treated as an image. The convolution and pooling operations are then applied to the mapped item embeddings.  
	In this paper, we first examine the typical session-based CNN recommender and show that  both the generative model and network architecture are suboptimal when modeling
	long-range dependencies  in the item sequence. 
	To address the issues, we introduce a simple, but very effective generative  model that is capable of learning high-level representation from both short- and long-range item dependencies.  
	The network architecture of the proposed model is formed of a stack of \emph{holed} convolutional layers, which can efficiently increase the receptive fields without relying on  
	the pooling operation.
	Another contribution is the effective use of residual block structure in recommender systems, which can ease the optimization for much deeper networks.
	The proposed generative model attains 
	state-of-the-art accuracy with less training time in the next item recommendation task.  
	It accordingly can be used as a powerful recommendation baseline to beat in future, especially when there are long sequences of user feedback.
\end{abstract}

%
%





\maketitle

\section{Introduction}
\label{Introduction}

Leveraging sequences of \revised{user-item} interactions (e.g., clicks or purchases) to improve real-world recommender systems has become increasingly popular in recent years. These sequences are automatically generated when users interact with online systems in sessions (e.g., shopping session, or music listening session).
For example, users on Last.fm\footnote{\scriptsize https://www.last.fm} or Weishi\footnote{\scriptsize http://weishi.qq.com/} typically enjoy a series of songs/videos during a certain time period without any interruptions, i.e., a listening or watching session. The set of music videos played in one session usually have strong correlations \cite{Zhiyong}, e.g., sharing the same album, writer, or genre. Accordingly, a good recommender system is supposed to generate recommendations by taking advantage of these sequential patterns in the session.

A class of models often employed for these sequences of interactions are the Recurrent Neural Networks (RNNs). RNNs typically generate a softmax output where high probabilities represent the most relevant recommendations. While effective,
these RNN-based models, such as \cite{hidasi2015session,Chatzis2017}, depend on a hidden state of the entire past that cannot fully utilize parallel
computation within a sequence ~\cite{gehring2017convolutional}. 
Thus their speed is limited in both training and evaluation. 

By contrast, training CNNs does not depend on the computations of the previous
time step and therefore allow parallelization over every element
in a sequence.  Inspired by the successful use of CNNs in image tasks,
\begin{figure}[h]
	\centering
	\small	
	\includegraphics[width=0.45\textwidth]{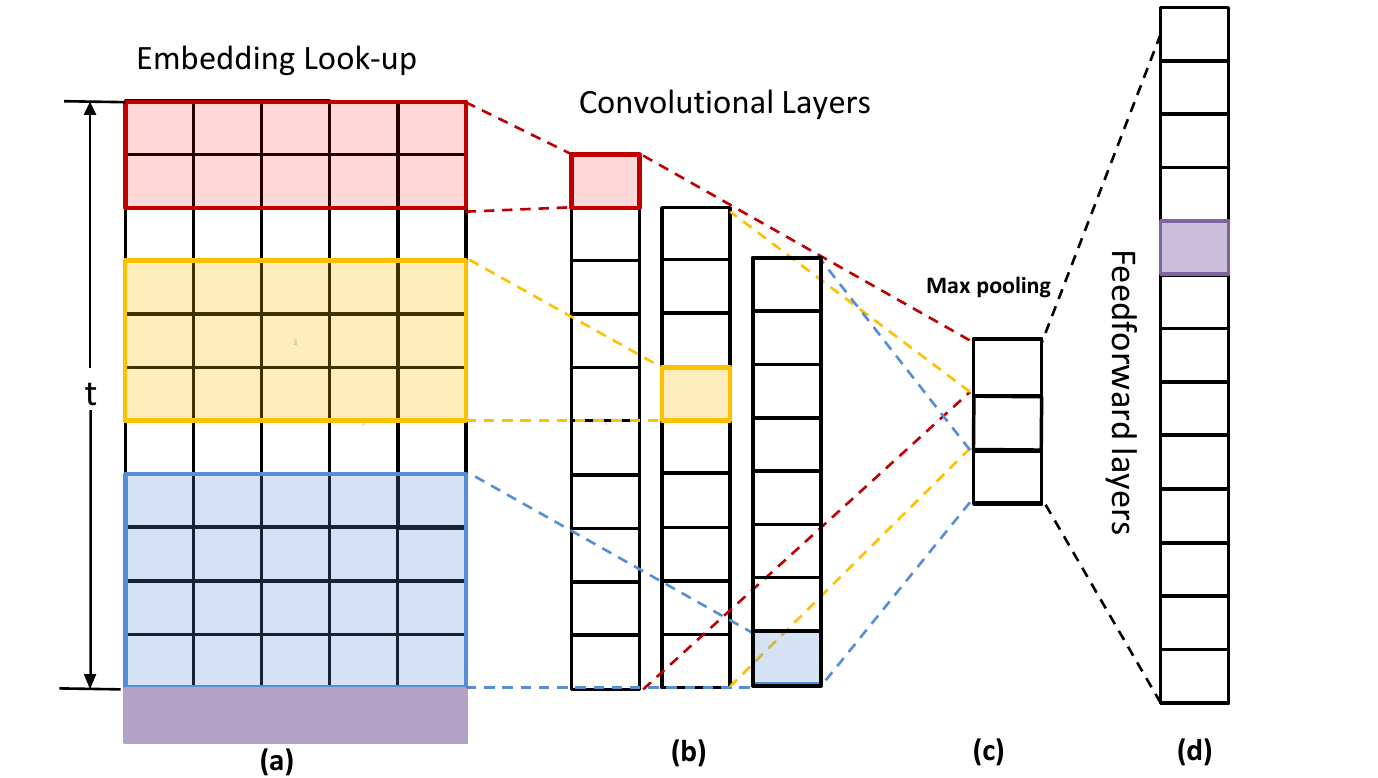}	
	\caption{\small The basic structure of \emph{Caser} \cite{tang2018caser}. The red, yellow and blue regions denotes a $2\times k$, $3\times k$ and $4\times k$ convolution filter respectively, where $k=5$. The purple row stands for the true next item.}
	\label{caser}
\end{figure}
a newly proposed sequential recommender, referred to as \emph{Caser} \cite{tang2018caser}, abandoned RNN structures, proposing instead a convolutional sequence embedding model, and demonstrated that
this CNN-based recommender is able to achieve comparable or superior performance to the popular RNN model in the top-$N$ sequential recommendation task. 
The basic idea of the convolution processing is to treat the  $t \times k$ embedding matrix as the ``image" of the previous $t$ \revised{interactions}  in  $k$ dimensional latent space and regard the sequential pattens as local features of  the ``image". A max pooling operation that only preserves the maximum value of the convolutional layer is performed to increase the receptive field, as well as dealing with the varying length of input sequences. Fig.~\ref{caser} depicts the key architecture of \emph{Caser}.

Considering the training speed of networks, in this paper we follow the path of sequential convolution techniques for the next item recommendation task. We show that  the typical network architecture used in \emph{Caser} has several obvious drawbacks --- e.g.,: (1) the max pooling scheme that is safely used in computer vision may discard important position and recurrent signals when modeling long-range sequence data; (2) generating the softmax distribution only for the desired item  fails to effectively use the compete set of dependencies. 
Both drawbacks become more severe
as the length of the sessions and sequences increases.
To address these issues, we introduce a simple but fundamentally  different CNN-based sequential recommendation model that 
allows us to model the complex conditional distributions even in very long-range item sequences.
To be more specific, first our generative model is designed to explicitly encode item inter-dependencies, which allows to directly estimates the distribution of the output sequence (rather than the desired item) over the raw item sequence.
Second, instead of using inefficient huge filters, we stack the 1D dilated convolutional layers  \cite{yu2015multi} on top of each other to increase the receptive fields when modeling long-range dependencies. The pooling layer can be safely removed in the proposed network structure.
It is worth noting that although the dilated convolution was invented for dense prediction in image generation tasks \cite{chen2016deeplab,yu2015multi,sercu2016dense}, and 
has been applied in other fields \revised{(e.g., acoustic \cite{oord2016wavenet,sercu2016dense}} and translation \cite{kalchbrenner2016neural} tasks), it is yet unexplored in recommender systems with huge sparse data. 
Furthermore, to ease the optimization of the deep generative architecture, \revised{we propose using residual network to 
	wrap  convolutional layer(s) by residual block.} 
To the best of our knowledge, this is \revised{also} the first work in terms of adopting residual learning to model the recommendation task.
The combination of these choices enables us to tackle large-scale problems and attain state-of-the-art results in both short- and long-range sequential recommendation data sets.
In summary, our main contributions include a novel \revised{recommendation} generative model (Section \ref{conditionmodel}) and a fundamentally different convolutional network architecture (Sections \ref{Network Architecture} $\sim$ \ref{Network Training}).   

\section{Preliminaries}
First, the problem of recommending items from sequences is described. Next, a recent convolutional sequence embedding recommendation model (\emph{Caser}) is shortly recapitulated along with its limitations. Lastly, we review previous work on sequence-based recommender systems.
\subsection{Top-$N$ Session-based Recommendation}
Let $\{x_0,x_1,...,x_{t-1},x_t\}$ (interchangeably denoted by $x_{0:t}$) be an \revised{user-item interaction sequence}  (or a session), where $x_i \in \mathbb{R}$ $(0 \leq  i \leq t)$ is the index of the clicked item out of a total number of $t+1$ items in the sequence.  
The goal of sequential recommendation is to seek a  model such that for a given prefix item sequence, $\textbf{x}=\{x_0,...,x_{i}\}$ $(0 \leq  i < t)$, it generates a ranking or classification distribution $\textbf{y}$ for all candidate items,
where $\textbf{y}=[ y_1,...,y_n ] \in \mathbb{R}^n $. $y_j$ can be a score, probability or a rank of item $i+1$ that will occur in this sequence. In practice, we typically make more than one recommendation by choosing the top-$N$ items from $\textbf{y}$, referred to as the top-$N$ session-based (sequential) recommendations.
\subsection{Limitations of \emph{Caser}}

The basic idea of \emph{Caser} is to embed the previous $t$ items as a $t\times k$ matrix $\bm{E}$ by the embedding look-up operation, as shown in Fig.~\ref{caser}~(a).
Each row vector of the matrix corresponds to the latent features of one item.
The embedding matrix can be regarded as the ``image" of the $t$ items in the $k$-dimensional latent space. Intuitively, models of various CNNs that are successfully applied in computer vision can be adapted to model the ``image" of an item sequence. However, there are two aspects that differentiate sequence modeling from image processing, which makes the use of CNN based models  non-straightforward.  First, the variable-length item sequences in real-world scenarios produce a large number of ``images" of different sizes, where traditional convolutional structures with fix-sized filters may fail. Second, the most effective filters for images, such as $3 \times 3$ and $5\times 5$, are not suitable for sequence ``images" since these small filters (in terms of row-wise orientation) are not suitable to capture the representations of full-width embedding vectors.

To address the above limitations, filters in \emph{Caser} slide over full columns of the sequence ``image'' by large filter. That is, the width of filters is usually the same as the width of the input ``images". The height typically varies by sliding windows over $2-5$ items at a time (Fig. \ref{caser} (a)). 
Filters of different sizes will generate variable-length feature maps after convolution (Fig. \ref{caser} (b)). To ensure that all maps have the same size,  the max pooling is performed over each map, which selects only the largest number of each feature map, resulting in a $1\times 1$ map (Fig. \ref{caser} (c)). 
Finally, these $1\times 1$ maps from all filters are concatenated to form a feature vector, followed by a softmax layer that yields the probabilities of next item  (Fig. \ref{caser} (d)). 
Note that we have omitted the vertical convolution in Fig. \ref{caser}, since it does not solve the major problems discussed below. 

Based on the above analysis of the convolutions in \emph{Caser}, one may find that there exist several drawbacks with the current design. First, the max pooling operator has obvious disadvantages. It cannot distinguish whether an important feature in the map occurs just one or multiple times and it ignores the position in which it occurs. The max pooling operator while safely used in image processing (with small pooling filters, e.g., $3\times 3$) may be harmful for modeling  long-range sequences (with large filters, e.g., $1 \times 20$).
Second, the shallow network structure in \emph{Caser} that suits for only one hidden convolutional layer is likely to fail when modeling complex relations or long-range dependences. 
The last important disadvantage comes from the generative process of next item, which we will describe  in detail in Section~\ref{conditionmodel}.
\subsection{Related Work}
\begin{figure*}[!t]
	\centering
	\subfloat[\scriptsize Standard CNN]{\includegraphics[width=0.31\textwidth]{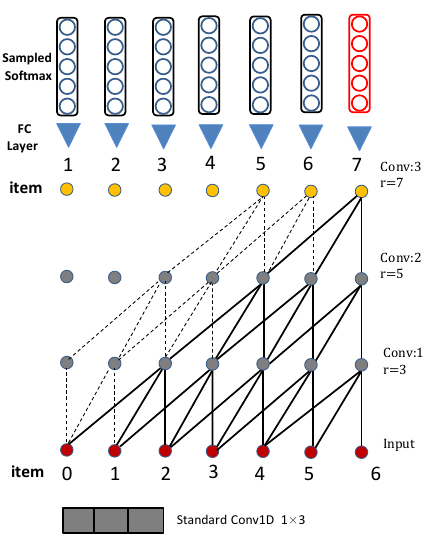}\label{vanallia}}
	\subfloat[\scriptsize Dilated CNN with `Holes']{\includegraphics[width=0.615\textwidth]{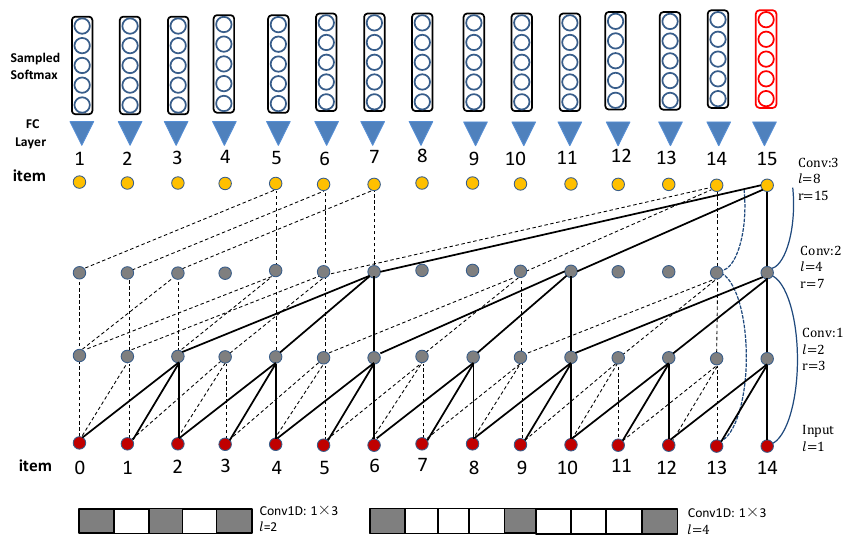}\label{dilated}}
	\caption{\small The proposed generative architecture with 1D standard CNNs  \text{(a)} and efficient dilated CNNs \text{(b)}. \revised{The blue  lines are the identity map which exists only for residual block (b) in Fig. \ref{residualreshape}}.
		An example of a standard 1D convolution filter and dilated filters are shown at the bottom of (a) and (b) respectively. We will refer to a dilated convolution with a dilation factor $l$ as $l$-dilated convolution.
		Apparently, compared with the standard CNN that linearly increases the receptive field by the depth of the network, the dilated CNN has a much larger receptive field by the same stacks without introducing more parameters. It can be seen that the standard convolution is a special case of 1-dilated convolution.}
	\label{generativestructure}
\end{figure*}
Early work in sequential recommendations mostly rely on the markov chain \cite{cheng2013you} and feature-based matrix factorization \cite{he2017neural,yuan2016lambdafm,yuan2017boostfm,yuan2018fbgd} approaches. Compared with neural network models, the markov chain based approaches fail to model complicated relations in the sequence data. For example, in \emph{Caser}, the authors showed that  markov chain approaches failed to model union-level sequential patterns and did not allow skip behaviors in the item sequences. Factorization based approaches such as factorization machines model a sequence by the sum of its item vectors.
However, these methods do not consider the order of items and are not specifically invented for sequential recommendations. 

Recently, deep learning models have shown state-of-the-art recommendation accuracy in contrast to conventional models. Moreover, RNNs, a class of deep neural networks, have almost dominated the area of sequential recommendations. For example, a Gated Recurrent Unit (\emph{GRURec}) architecture with a ranking loss was proposed by \cite{hidasi2015session} for session-based recommendation. In the follow-up papers,  various RNN variants have been designed to extend the typical one for different application scenarios, such as by adding personalization \cite{quadrana2017personalizing}, content~\cite{gu2016learning} and contextual features~\cite{Elena},  attention mechanism \cite{CuiQiang,li2017neural} and different ranking loss functions~\cite{hidasi2017recurrent}. 

By contrast, CNN based sequential recommendation models are more challenging and much less explored because convolutions are not a natural way to capture sequential patterns.  To our best knowledge, only two types of sequential recommendation architectures have been proposed to date: the first one by \emph{Caser} is a standard 2D CNN, while the second is a 3D CNN~\cite{Tuan:2017:CNS:3109859.3109900} designed to model high-dimensional features. Unlike the aforementioned examples, we plan to investigate the effects of 1D CNNs with efficient dilated convolution filters and residual blocks for building the recommendation architecture.

\section{Model Design}
To address the above limitations, we introduce a new probabilistic generative model that is formed of a stack of 1D convolution layers. We first focus on the form of the distribution, and then the architectural innovations.
Generally, our proposed model is fundamentally different from \emph{Caser} in several key ways: (1) our probability estimator explicitly models the  distribution transition of all individual items  at once, rather than the final one,  in the sequence;
(2) our  network has a deep, rather than shallow, structure; (3) our convolutional layers are based on the efficient 1D dilated convolution rather than standard 2D  convolution; and (4) pooling layers are removed. 

\subsection{A Simple Generative Model}
\label{conditionmodel}
In this section, we introduce a simple yet very effective generative model directly operating on the sequence of previous interacted items. 
Our aim is to estimate a distribution over the original item interaction sequences that can be  used to \emph{tractably} compute the likelihood of the items and to generate the future items that users would like to interact.
Let  $p(\textbf{x})$ be the joint distribution of item sequence $\textbf{x}=\{x_0, ...,x_{t}\}$. To model  $p(\textbf{x})$, we can factorize it as a product of
conditional distributions by the chain rule. 
\begin{equation}
	\label{jointdis}
		p(\textbf{x})=\prod_{i=1}^{t}p(x_i|x_{0:{i-1}},\bm{\theta})p(x_0)
\end{equation}
where the value $p(x_i|x_{0:{i-1}},\bm{\theta})$ is the probability of $i$-th item $x_i$ conditioned on all the previous items $x_{0:{i-1}}$. 
A similar setup has been explored by  NADE \cite{larochelle2011neural}, PixelRNN/CNN \cite{oord2016pixel,oord2016conditional} in biological and image  domains.


Owing to the ability of neural networks in modeling complex nonlinear relations, in this paper we model the conditional distributions of user-item interactions by a stack of 1D convolutional networks. To be more specific, the network receives $x_{0:{t-1}} $ as the input and outputs distributions over possible $x_{1:{t}}$, where the distribution of $x_t$ is our final expectation. For example, as illustrated in Fig. \ref{generativestructure}, the output distribution of $x_{15}$ is determined by $x_{0:14}$, while $x_{14}$is determined by $x_{0:13}$. 
It is worth noting that in previous sequential recommendation literatures, such as \emph{Caser}, \emph{GRURec} and \cite{li2017neural,quadrana2017personalizing,Tuan:2017:CNS:3109859.3109900,tan2016improved},  they only model a single conditional distribution $p(x_i|x_{0:{i-1}},\bm{\theta})$ rather than all conditional probabilities $\prod_{i=1}^{t}p(x_i|x_{0:{i-1}},\bm{\theta})p(x_0)$. Within the context of the above example, assuming $\{x_0, ...,x_{14}\}$ is given, models like \emph{Caser} only estimate the probability distribution (i.e.,  softmax) of the next item $x_{15}$ (also see Fig. \ref{caser} (d)), while our generative method estimates the distributions 
of all individual items in $\{x_1, ...,x_{15}\}$. The comparison of the generating process is shown below.
\begin{equation}
\label{subsess}
\begin{aligned}
&Caser/GRURec: \underbrace{\{x_0, x_1,...,x_{14}\}}_{input}\Rightarrow \underbrace{x_{15}}_{output}\\
&  Ours: \underbrace{ \{x_0, x_1,...,x_{14}\}}_{input}\Rightarrow  \underbrace{\{x_1, x_2,...,x_{15}\}}_{output}
\end{aligned}
\end{equation}
where $\Rightarrow$ denotes `predict'. Clearly, our proposed model is more effective in capturing the set of all sequence relations,
whereas \emph{Caser} and \emph{GRURec} fail to \emph{explicitly} model the internal sequence features between $\{x_0, ...,x_{14}\}$.
In practice, to address the drawback, such models will typically generate a number of sub-sequences (or sub-sessions) for training by means of data augmentation techniques~\cite{tan2016improved} (e.g., padding, splitting or shifting the input sequence), such as shown in Eq.~(\ref{createsubsess}) ( see~\cite{Tuan:2017:CNS:3109859.3109900,li2017neural,quadrana2017personalizing,tang2018caser}).

\begin{equation}
\label{createsubsess}
\begin{aligned}
& Caser/GRURec \ sub-session-1: \{x_{-1}, x_0,...,x_{13}\}\Rightarrow x_{14}\\
&  Caser/GRURec \ sub-session-2: \{x_{-1}, x_{-1},...,x_{12}\}\Rightarrow x_{13}\\
&\qquad \qquad \qquad \qquad\qquad......\\
&  Caser/GRURec \ sub-session-12: \{x_{-1}, x_{-1},...,x_{2}\}\Rightarrow x_{3}\\
\end{aligned}
\end{equation}
While effective, the above approach to generate sub-session cannot guarantee the optimal results due to the separate optimization for each sub-session. In addition, optimizing these sub-sessions separately will result in  corresponding computational costs. Detailed comparison with empirical results has also been reported in our experimental sections.


\subsection{Network Architecture}
\label{Network Architecture}

\subsubsection{Embedding Look-up Layer:}
 Given an item sequence $\{x_0,...,x_{t}\}$, the model retrieves each of the first $t$ items $\{x_0,...,x_{t-1}\}$ via a look-up table, 
and stacks these item embeddings together. Assuming the embedding dimension is $2k$, where $k$ can be set as the number of inner channels in the convolutional network. This results in a  matrix of size $t \times 2k$. Note that unlike \emph{Caser} that treats the input matrix as a 2D ``image" during convolution, our proposed architecture learns the embedding layer by 1D  convolutional filters, which we will describe later.
\subsubsection{Dilated layer:}
  As shown in Fig. \ref{generativestructure} (a), the standard filter is only able to perform convolution with the receptive field linearly 
   by the depth of the network. This makes it difficult to 
  handle long-range sequences.
 Similar to Wavenet \cite{oord2016wavenet}, we employ the dilated convolution  to construct the proposed generative model. 
 The basic idea of dilation is to
apply the convolutional filter over a field larger 
than its original length by dilating it with zeros. As such, it is more efficient since it utilizes fewer parameters. For this reason, a dilated filter is also referred to as a \emph{holed} or \emph{sparse} filter. Another benefit is that dilated convolution can preserve the spatial dimensions of the input, which makes the stacking operation much easier for both convolutional layers and residual structures. 

Fig. \ref{generativestructure} shows the network comparison between the standard convolution and dilated convolutions with the proposed sequential generative model. 
The dilation factor in (b) are $1,2,4$ and $8$. 
To describe the network architecture, we denote  receptive field, $j$-th convolutional layer, channel and dilation as $r$, $F_j$, $C$ and $l$ respectively. By setting the width of convolutional filter $f$ as 3, we can see that the  dilated convolutions (Fig. \ref{generativestructure} (b)) allow for exponential increase in the size of receptive fields ($r=2^{j+1}-1$), while the same stacking structure for the standard convolution  (Fig. \ref{generativestructure} (a)) has only linear
receptive fields ($r=2j+1$).  
Formally, with dilation $l$, the filter window from location $i$ is given as
\begin{equation*}
\left[x_i \quad x_{i+l}\quad x_{i+2l}\quad...\quad x_{i+(f-1)\cdot l} \right]
\end{equation*}
and the 1D dilated convolution operator $*_{l}$ on element $h$ of the item sequence is given below
\begin{equation}
(\textbf{x}*_{l}g)(h)=\sum_{i=0}^{f-1} \textbf{x}_{h-l\cdot i}\cdot g(i)
\end{equation}
where $g$ is the filter function.
Clearly, the dilated convolutional structure is more effective to model long-range item sequences, and thus more efficient without using larger filters or becoming deeper.   
In practice, to further increase the model capacity and receptive fields, one just need to repeat the architecture in Fig.~\ref{generativestructure} multiple times by stacking, e.g., $1,2,4,8, 1,2,4,8$. 



\subsubsection{One-dimensional Transformation:}
\begin{figure}[!t]
	\centering
	\subfloat[\scriptsize ]{\includegraphics[width=0.1\textwidth]{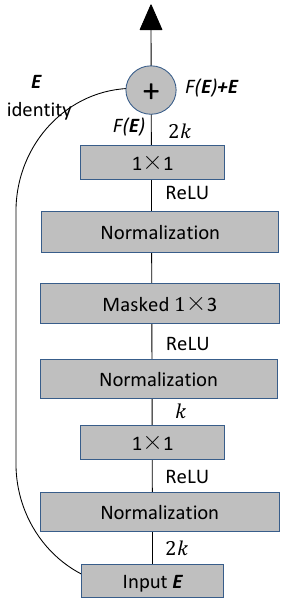}\label{vanallia}}
	\subfloat[\scriptsize ]{\includegraphics[width=0.1\textwidth]{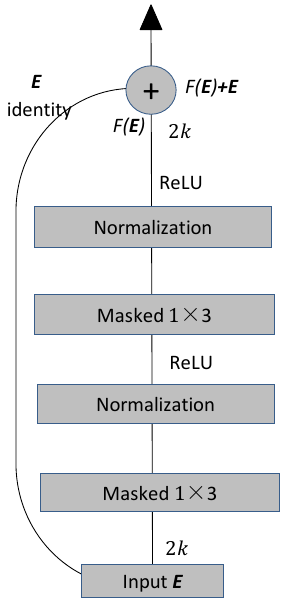}\label{vanallia2}}
	\subfloat[\scriptsize One-dimensional Transformation]{\includegraphics[width=0.23\textwidth]{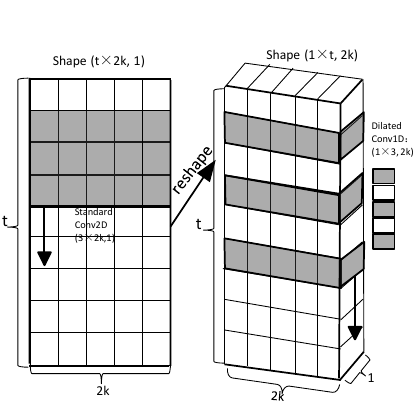}\label{dilated}}
	\caption{\small Dilated residual blocks (a), (b) and one-dimensional transformation (c). (c) shows the transformation from the 2D filter ($C= 1$)(left) to the 1D 2-dilated filter ($C= 2k$) (right); the vertical black arrows represent the  direction of the sliding convolution. In this work, the default stride for the dilated convolution is~1. 
			Note the reshape operation in (b) is performed before each convolution in (a) and (b) (i.e., $1\times1$ and masked $1\times 3$), which is then followed by a reshape back step after convolution. }
	\label{residualreshape}
\end{figure}

Although our dilated convolution operator depends on the 2D input matrix $\bm{E}$, the proposed network architecture is actually composed of all 1D convolutional layers. To model the 2D embedding input, we perform a simple reshape operation, which serves as a prerequisite for performing 1D convolution. Specifically, the 2D matrix $\bm{E}$ is reshaped from $t\times 2k$ to a 3D tensor $\bm{T}$ of size $1\times t \times 2k$, where $2k$ is treated as the ``image'' channel rather than the width of the standard convolution filter in \emph{Caser}.  
Fig. \ref{residualreshape}  (b) illustrates the reshaping process. 


\begin{figure*}[!t]
	\centering
	\small	
	\includegraphics[width=0.99\textwidth]{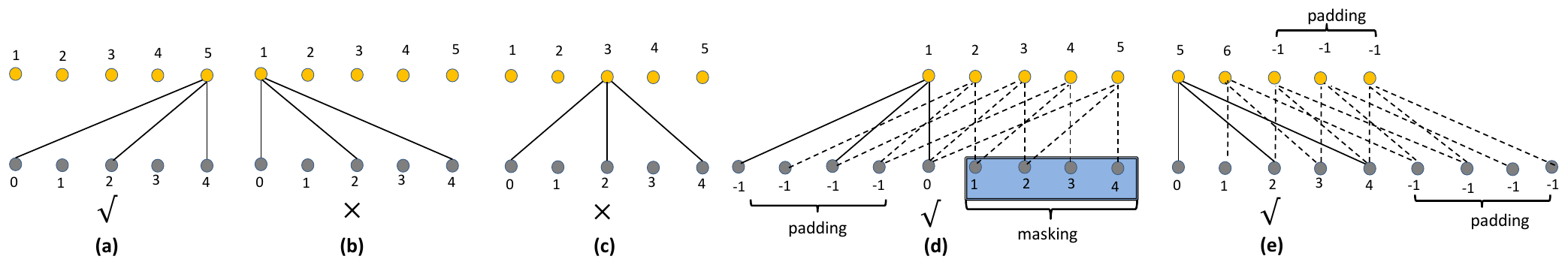}	
	\caption{\small The future item can only be determined by the past ones according to Eq. (\ref{jointdis}).  (a) (d) and (e) show the correct convolution process, while (b) and (c) are wrong. E.g., in (d), items of $\{1,2,3,4\}$ are masked when predicting $1$, which can be technically implemented by padding.}
	\label{mask}
\end{figure*} 

\subsection{Masked Convolutional  Residual Network}
\label{ResidualLearning}
Although increasing the depth of network layers can help obtain higher-level feature representations, it also easily results in the vanishing gradient issue, which makes the learning process much harder. To address the degradation problem, residual learning \cite{he2016deep} has been introduced for deep networks. While residual learning has achieved huge success in the domain of computer vision, it has not appeared in the recommender system literature.  

The basic idea of residual learning is to stack multiple convolutional layers together as a block and then employ a skip connection scheme that passes the previous layers's feature information to its posterior layer.  The  skip connection scheme allows to explicitly fit the  residual mapping rather than the original identity mapping, which can maintain the input information and thus enlarge the propagated gradients. Formally, denoting the  desired mapping as $H(\bm{E})$, we let the residual block fit another mapping of $F(\bm{E})=H(\bm{E})-\bm{E}$. The desired mapping now is recast into $F(\bm{E})+\bm{E}$ by element-wise addition (assuming that $F(\bm{E})$ and $\bm{E}$ are of the same dimension).  As has been evidenced in \cite{he2016deep}, optimizing  the residual mapping $F(\bm{E})$ is much easier than the original, unreferenced mapping $H(\bm{E})$.  \revised{Inspired by \cite{he2016identity,kalchbrenner2016neural}, we introduce two residual modules in Fig. \ref{residualreshape}~(a) and (b)}.


In (a), we wrap each dilated convolutional layer by a residual block, while in (b)  we wrap every two dilated layers by a  different residual block.
	That is, with the design of block (b), the input layer and the second convolutional layer should be connected by skip connection (i.e., the blue lines in Fig. \ref{generativestructure}).	
	Specifically, 
	each block is made up of the normalization, activation (e.g., ReLU \cite{nair2010rectified}), convolutional layers and a skip connection  in a specific order.  In this work we adopt the state-of-the-art layer normalization~\cite{ba2016layer} before each activation layer, as it is  well suited to sequence processing and online learning in contrast with batch normalization~\cite{ioffe2015batch}.

Regarding the properties of the two residual networks, the residual block in (a) consists of 3 convolution filters: one dilated filter of size $1\times 3 $ and two regular filters of size $1\times1$. The  $1\times1$ filters are introduced to change the size of  $C$ so as to reduce the parameters to be learned by the $1\times 3$ kernel. The  first $1\times1$ filter (close to input $\bm{E}$ in Fig. \ref{residualreshape} (a)) is to change $C$ from $2k$ to $k$, while the second  $1\times1$ filter does the opposite transformation in order to maintain the spatial dimensions for the next stacking operation. To show the effectiveness of the   $1\times1$ filters in (a), we compute the number of parameters in both (a) and (b).  For simplicity, we omit the activation and normalization layers. 
	As we can see, the number of parameters for the $1\times 3 $ filter is $1\times 3 \times 2k \times 2k= 12k^2$  (i.e., in (b)) without the  $1\times1$ filters. While in (a),  the number of parameters to be learned is $1\times 1\times 2k\times k +1\times 3\times k\times k+ 1\times 1 \times k \times 2k= 7k^2$. 
	The residual mapping $F(\bm{E},\{W_i\})$ in (a) and (b) is formulated as:
\begin{equation}
\label{residualmapping}
F(\bm{E},\{W_i\})=\begin{cases}	W_3(\sigma(\psi(W_2(\sigma(\psi(W_1(\sigma(\psi(\bm{E}) )))))))))

& \text{Fig.} \ref{residualreshape} \ \text{(a)} \\\sigma(\psi(W'_4(\sigma(\psi(W'_2(\bm{E})))))                & \text{Fig.} \ref{residualreshape}\ \text{(b)}  \end{cases}
\end{equation}
where $\sigma$ and $\psi$ denote ReLU and layer-normalization, $W_1$ and $W_3$ denote the convolution \revised{weight function} of standard $1\times 1$ convolutions, and $W_2$, $W'_2$ and $W'_4$ denote the \revised{weight function}  of $l$-dilated convolution filter with size of  $1\times 3$. Note that bias terms are omitted for simplifying notations.

\subsubsection{Dropout-mask:}
To avoid the future information leakage problem, we propose a masking-based dropout trick for the 1D dilated convolution to prevent the network from  seeing the future items.
Specifically, when predicting $p(x_i|x_{0:{i-1}})$, the convolution filters are not allowed to make use of the information from $x_{i:t}$. Fig. \ref{mask} shows several different ways to perform convolution. As shown, our dropout-masking operation can be implemented either by  padding the input sequence in (d) or shifting the output sequence by a few timesteps in (e). The padding method in (e) is very likely to result in information loss in a sequence, particularly for short sequences. Hence in this work, we apply the padding strategy in (d) with the padding size of  $(f-1)*l$.

\subsection{Final Layer, Network Training and Generating}
\label{Network Training}

As mentioned, the  matrix in the last layer   of the convolution architecture (see Fig. \ref{generativestructure}), denoted by $ \bm{E}^{o}$ , preserves the same dimensional size of the input  $\bm{E}$, i.e., $\bm{E}^{o} \in \mathbb{R}^{t\times 2k}$. However, the output should be a matrix or tensor that contains probability distributions of all items in the output sequence 
$x_{1:t}$, where the probability distribution of $x_t$ is the desired one that generates top-$N$ predictions. To do this, \revised{we can simply use one more} convolutional layer on top of the last convolutional layer in Fig. \ref{generativestructure} with filter of size $1\times 1 \times 2k \times n$, where $n$ is the number of items. Following the procedure of one-dimensional transformation  in Fig. \ref{residualreshape}  \revised{(c)}, we  obtain the expected ouput matrix $\bm{ E}^p \in \mathbb{R}^{t\times n}$, where each row vector after the softmax operation represents the categorical distribution over $x_i$ $(0 < i \leq t)$. 

The aim of  optimization is to maximize the log-likelihood of the training data w.r.t.  $\bm{\theta}$. Clearly, maximizing $\log p(\textbf{x}) $  is mathematically  equivalent to minimizing the \emph {sum} of the binary cross-entropy loss for each item in $x_{1:t}$. For practical recommender systems with tens of millions items, the negative sampling strategy can be applied to bypasses the generation of full softmax distributions, where the $1 \times 1$ convolutional layer is replaced by a fully-connected (FC) layer with weight matrix $\bm{ E}^g \in \mathbb{R}^{2k\times n}$. For example, we can apply either the sampled softmax \cite{jean2014using} or kernel based sampling~\cite{blanc2017adaptive}. The recommendation accuracy by these negative sampling strategies is nearly identical with the full softmax method with properly tuned sampling size.

For  comparison purpose,  we only predict the next one item in our evaluation, and then stop the generating process. Nevertheless, the model is able to generate a sequence of items simply by feeding the predicted one item (or sequence)  into the network to predict the next one, and thus the prediction at the generating phrase is \emph {sequential}. 
This matches most real-world recommendation scenarios, where the next action is followed when the current one has been observed.
But at both training and evaluation phases, the conditional predictions for all timesteps can be made \emph {in parallel}, because the complete sequence of input items  $\textbf{x}$ is already available.

\begin{table*} 
	\centering
	\caption{\small Session statistics of  all data sets. }
	\small
	\label{sessionsta}
	\begin{threeparttable}				
		\begin{tabular}{@{} L{21mm} C{16mm} C{16mm} C{16mm} C{16mm}C{16mm}C{17mm}C{19mm} @{}}
			\toprule
			\small DATA &  \small \emph{YOO} &  \small \emph{ MUSIC\_M5 }&  \small \emph{MUSIC\_L5} &  \small \emph{MUSIC\_L10}&  \small \emph{MUSIC\_L20}&  \small \emph{MUSIC\_L50}&  \small \emph{MUSIC\_L100} \\
			\midrule
			\midrule
			\emph{RAW-SESSIONS}          &   0.14M& 0.61M & 2.14M & 1.07M& 0.53M& 0.21M& 0.11M  \\ 	
			\emph{SUB-SESSIONS-T}           &   0.07M& 0.31M & 1.07M & 3.21M&4.28M& 4.91M& 5.10M  \\
			
			\bottomrule
		\end{tabular}
		\scriptsize \emph{MUSIC\_M5} denotes \emph{MUSIC\_M} with maximum session size of $5$. The same applies to \emph{MUSIC\_L}. `M' denotes 1 million. 
	\end{threeparttable}
\end{table*}


\section{Experiments}
\label{EXPERIMENTS}
In this section we detail our experiments, report results for several data sets, and compare our model (called \emph{NextItNet}) with the well-known RNN-based  model \emph{GRURec} \cite{hidasi2015session, tan2016improved} and the state-of-the-art CNN-based model \emph{Caser}. 
 Note that (1) since the main contributions  in this paper do not focus on combining various features, we omit the comparison with content- or context-based sequential recommendation models, such as the 3D CNN recommender \cite{Tuan:2017:CNS:3109859.3109900} and other RNN variants \cite{gu2016learning,Elena,li2017neural,quadrana2017personalizing};  (2) the \emph{GRURec} baseline could be  regarded as the state-of-the-art \emph{Improved GRURec} \cite{tan2016improved} when dealing with the long-range session data sets because our main data augmentation technique for the two baseline models follows the same way in \emph{Improved GRURec}.
\subsection{Datasets and Experiment Setup}
\subsubsection{Datasets and Preprocessing}
The first data set `Yoochoose-buys' (YOO for short) is chosen from the RecSys Challenge 2015\footnote{\scriptsize http://2015.recsyschallenge.com/challenge.html}.  We use the buying dataset for evaluation.
For preprocessing, we filter out sessions of length shorter than $3$. 
Meanwhile, we find that in the processed Yoo data 96$\%$ sessions have a length shorter than $10$, and we simply remove the 4$\%$ longer sessions and refer it as a short-range sequential data.

The remaining data sets are extracted from Last.fm\footnote{\scriptsize http://www.dtic.upf.edu/~ocelma/MusicRecommendationDataset/lastfm-1K.html}: one medium-size (\emph {MUSIC\_M}) and one large-scale (\emph {MUSIC\_L}) collection by randomly drawing 20,000 and 200,000 songs respectively. In the Last.fm data set, we observe that most users listen to music several hundred times a week, and some even listen to more than one hundred songs within a day. Hence, we are able to test our model in both short- and long-range sequences by cutting up these long-range listening sessions. In \emph {MUSIC\_L},  we define the maximum session length $t$ as $5$, $10$, $20$, $50$ and $100$, and then extract every $t$ successive items as our input sequences. This is done by sliding a window of both size and stride of $t$ over the whole data. We ignore sessions in which the time span between the last two items is longer than 2 hours.
In this way, we create  5 data sets, referred to as \emph{RAW-SESSIONS}. We randomly split these \emph{RAW-SESSIONS} data into training (50\%), validation (5\%), and testing (45\%) sets.

\begin{table*} 
	\centering
	\caption{\small Accuracy comparison. The upper, middle and below tables are MRR@$5$, HR@$5$  and NDCG@$5$ respectively. }
	\small
	\label{overallresults5}
	\begin{threeparttable}				
		\begin{tabular}{@{} L{16mm} C{17mm} C{17mm} C{17mm} C{17mm}C{17mm}C{17mm}C{19mm} @{}}
			\toprule
			\small DATA &  \small \emph{YOO} &  \small \emph{ MUSIC\_M5 }&  \small \emph{MUSIC\_L5} &  \small \emph{MUSIC\_L10}&  \small \emph{MUSIC\_L20}&  \small \emph{MUSIC\_L50}&  \small \emph{MUSIC\_L100} \\
			\midrule
			\midrule
			\emph{MostPop}          &   0.0050& 0.0024 & 0.0006 & 0.0007&0.0008&0.0007& 0.0007  \\ 
			\emph{GRURec}           &   0.1645& 0.3019 & 0.2184 & 0.2124& 0.2327 &0.2067& 0.2086 \\ 
			\emph{Caser}              &  0.1523& 0.2920&0.2207 & 0.2214& 0.1947&0.2060& 0.2080  \\ 
			\emph{NextItNet}           &   \textbf{0.1715}& \textbf{0.3133} & \textbf{0.2327}  &\textbf{ 0.2596} &\textbf{ 0.2748}&\textbf{0.2735} & \textbf{0.2583}  \\
			\midrule
			\emph{MostPop}         &  0.0151& 0.0054& 0.0014 & 0.0016 &0.0016& 0.0016& 0.0016  \\ 
			\emph{GRURec}           &  0.2773& 0.3610 &  0.2626&0.2660&0.2694& 0.2589& 0.2593  \\ 
			\emph{Caser}           &  0.2389&0.3368 & 0.2443&0.2631& 0.2433&0.2572&0.2588  \\ 
			\emph{NextItNet}          &  \textbf{0.2871}& \textbf{0.3754} &\textbf{0.2695} &\textbf{0.3014}& \textbf{0.3166} &\textbf{0.3218} &  \textbf{0.3067} \\
				\midrule
				\emph{MostPop}         &  0.0075& 0.0031&0.0008 & 0.0009 &0.0010& 0.0009& 0.0009  \\ 
				\emph{GRURec}           &  0.1923& 0.3166 & 0.2294 &0.2258&0.2419& 0.2197&0.2212 \\ 
				\emph{Caser}           &  0.1738&0.3032&0.2267&0.2318& 0.2068&0.2188& 0.2207 \\ 
				\emph{NextItNet}        &  \textbf{0.2001}& \textbf{ 0.3288} &\textbf{0.2419} &\textbf{0.2700}& \textbf{ 0.2853} &\textbf{0.2855} &  \textbf{0.2704} \\
			
			\bottomrule
		\end{tabular}
		\scriptsize \emph{MostPop} returns the most popular item respectively. Regarding the setup of our model, we use two-hidden-layer  convolution structure with dilation factor  $1,2,4$ for the first four data sets (i.e., \emph{YOO}, \emph{ MUSIC\_M5 }, \emph{MUSIC\_L5} and  \emph{MUSIC\_L10}), while for the last three long-range sequence data sets, we use $1,2,4,8, 1,2,4,8, $ to obtain above results.
	\end{threeparttable}
\end{table*}

\begin{table*} 
	\centering
	\caption{\small Accuracy comparison. The upper, middle and below tables are MRR@$20$, HR@$20$  and NDCG@$20$ respectively. }
	\small
	\label{overallresults20}
	\begin{threeparttable}				
		\begin{tabular}{@{} L{16mm} C{17mm} C{17mm} C{17mm} C{17mm}C{17mm}C{17mm}C{19mm} @{}}
			\toprule
			\small DATA &  \small \emph{YOO} &  \small \emph{ MUSIC\_M5 }&  \small \emph{MUSIC\_L5} &  \small \emph{MUSIC\_L10}&  \small \emph{MUSIC\_L20}&  \small \emph{MUSIC\_L50}&  \small \emph{MUSIC\_L100} \\
			\midrule
			\midrule
			\emph{MostPop}          &   0.0090&  0.0036 &0.0009 & 0.0010&0.0011&0.0011&0.0011  \\ 
			\emph{GRURec}           &   0.1839& 0.3103 & 0.2242 & 0.2203& 0.2374& 0.2151&  0.2162 \\ 
			\emph{Caser}              &  0.1660& 0.2979 &0.2234 & 0.2268& 0.2017&0.2133& 0.2153 \\ 
			\emph{NextItNet}           &   \textbf{0.1901}& \textbf{0.3223} & \textbf{0.2375}  &\textbf{0.2669} &\textbf{ 0.2815}&\textbf{0.2794} & \textbf{0.2650}  \\
			\midrule
			\emph{MostPop}         &  0.0590& 0.0180 &0.0052 & 0.0053 &0.0056& 0.0056& 0.0056  \\ 
			\emph{GRURec}           &   0.4603&0.4435& 0.3197&0.3434&0.3158& 0.3406&0.3336  \\ 
			\emph{Caser}           &  0.3714& 0.3937 & 0.2703&0.3150& 0.3110&0.3273&0.3298 \\ 
			\emph{NextItNet}          &  \textbf{0.4645}& \textbf{0.4626} &\textbf{0.3159} &\textbf{0.3709}& \textbf{0.3814} &\textbf{0.3789} &  \textbf{0.3731} \\
			\midrule
			\emph{MostPop}         &  0.0195& 0.0066&0.0018 & 0.0019 &0.0021& 0.0020& 0.0020  \\ 
			\emph{GRURec}           &  0.2460&0.3405 & 0.2460 &0.2481&0.2553&0.2433&0.2427\\ 
			\emph{Caser}           &  0.2122& 0.3197&0.2342& 0.2469& 0.2265&0.2392& 0.2412 \\ 
			\emph{NextItNet}        &  \textbf{0.2519}& \textbf{ 0.3542} &\textbf{0.2554} &\textbf{0.2904}& \textbf{0.3041} &\textbf{0.3021} &  \textbf{0.2895} \\
			
			\bottomrule
		\end{tabular}
	\end{threeparttable}
\end{table*}

\begin{table} 
	\centering
	\caption{\small Effects of sub-session in terms of MRR@5. 
		The upper, middle and below tables represent  GRU, Caser and NextItNet  respectively. ``10'',``20'',``50'' and ``100'' are the session lengths.	}
	\small
	\label{withsubsession}
	\begin{threeparttable}				
		\begin{tabular}{@{} L{18mm} C{12mm} C{12mm} C{12mm} C{12mm} @{}}
			\toprule
			\small Sub-session   &  \small 10 &  \small 20 &  \small 50 &  \small 100   \\
			\midrule
			\midrule

			\emph{Without}          &  0.1985&  0.1645 &0.1185 &0.0746 \\ 
			\emph{With}           &  0.2124& 0.2327& 0.2067 & 0.2086\\
			\midrule
			\emph{Without}          &  0.1571&  0.1012 &0.0216 &  0.0084 \\ 
			\emph{With}           &  0.2214& 0.1947 &0.2060 & 0.2080 \\

			\midrule
			\emph{Without}          &  0.2596& 0.2748&0.2735 &0.2583 \\ 		
			\bottomrule
		\end{tabular}
	\end{threeparttable}
\end{table}

\begin{table} 
	\centering
	\caption{\small Effects of the residual block in terms of MRR@5. ``\emph{Without}'' means no skip connection.
		``\emph{M5}'', ``\emph{L5}'', ``\emph{L10}'' and ``\emph{L50}'' denote  \emph{ MUSIC\_M5 }, \emph{ MUSIC\_L5 }, \emph{ MUSIC\_L10 } and \emph{ MUSIC\_L50  } respectively.		}
	\small
	\label{residual}
	\begin{threeparttable}				
		\begin{tabular}{@{} L{18mm} C{13mm} C{13mm} C{13mm} C{13mm} @{}}
			\toprule
			\small DATA   &  \small \emph{M5} &  \small \emph{L5} &  \small \emph{L10}&  \small \emph{L50}   \\
			\midrule
			\midrule

			\emph{Without}          &  0.2968&  0.2146 &0.2292 & 0.2432\\ 
			\emph{With}           &  0.3300& 0.2455& 0.2645 & 0.2760\\
			
			\bottomrule
		\end{tabular}
	\end{threeparttable}
\end{table}

As mentioned before, the performance of \emph{Caser} and \emph{GRURec} is supposed to degrade significantly for long sequence inputs, such as when $t=20$, $50$ and $100$. For example, when setting $t=50$, \emph{Caser} and \emph{GRURec} will predict $x_{49}$ by using $x_{0:48}$, but without explicitly modeling  the item inter-dependencies  between $x_0$ and $x_{48}$. To remedy this defect, when $t>5$, we follow the common approach \cite{tan2016improved,li2017neural} by manually creating additional sessions from the \emph{training}  sets of \emph{RAW-SESSIONS} so that \emph{Caser} and \emph{GRURec} can leverage the full dependency to a large extent. Still setting $t=50$, one training session will then produce $45$ more sub-sessions by padding the beginning and removing the end indices, referred to as \emph{SUB-SESSIONS}.  
The example of the $45$ sub-sessions are given as follows: $\{x_{-1}, x_0, x_1,...,x_{48}\}$, $\{x_{-1}, x_{-1}, x_0,...,x_{47}\}$,..., $\{x_{-1}, x_{-1}, x_{-1},...,x_{4}\}$. Regarding \emph {MUSIC\_M}, we only show the results when $t=5$. 
We show the statistics of \emph{RAW-SESSIONS} $\&$  training data of \emph{SUB-SESSIONS} (i.e., \emph{SUB-SESSIONS-T}) in Table \ref{sessionsta}. 

All models were trained on GPUs (TITAN V) using Tensorflow.  The learning rates and batch sizes of baseline methods were manually set according to performance in validation sets. For all data sets, \emph{NextItNet} used the learning rate of 0.001 and batch size of 32. Embedding size $2k$ is set to 64 for all models without special mention.   We report results with residual block (a) and full softmax. We have validated the performance of results block (b) separately. To further evaluate the effectiveness of the two residual blocks, we have also tested our model in another dataset, namely, Weishi\footnote{\scriptsize We are working to release this dataset, which has very good sequential property.}. The improvements are about two times compared with the same model without residual blocks.

\begin{table} 
	\centering
	\caption{\small Effects (MRR@$5$) of increasing embedding size. The upper and below tables are \emph{MUSIC\_M5 }  and \emph{MUSIC\_L100 }  respectively. }
	\small
	\label{embedding}
	\begin{threeparttable}				
		\begin{tabular}{@{} L{18mm} C{12mm} C{12mm} C{12mm} C{12mm} @{}}
			\toprule
			\small $2k$   &  \small 16 &  \small 32 &  \small 64 &  \small 128   \\
			\midrule
			\midrule
				\emph{GRURec}           &   0.2786&0.2955 &0.3019 &0.3001 \\ 
				\emph{Caser}            &  \textbf{0.2855}& 0.2982 & 0.2979 &0.2958  \\  
				\emph{NextItNet}            & 0.2793& \textbf{0.3063} & \textbf{0.3133} & \textbf{ 0.3183}  \\

			\midrule   	
					\emph{GRURec}           &  0.1523&  0.1826& 0.2086 &0.2043 \\ 
					\emph{Caser}            &  0.0643&0.1129& 0.2080 & 0.2339 \\  
					\emph{NextItNet}           &   \textbf{0.1668}& \textbf{0.2289} & \textbf{0.2583} & \textbf{ 0.2520} \\

			\bottomrule
		\end{tabular}
	\end{threeparttable}
\end{table}

\subsubsection{Evaluation  Protocols}
We reported the evaluated results by three popular top-$N$ metrics, namely MRR@$N$ (Mean Reciprocal Rank) \cite{hidasi2015session},  HR@$N$ (Hit Ratio) \cite{he2018adversarial} and NDCG@$N$ \cite{yuan2018fbgd} (Normalized Discounted Cumulative Gain). 
$N$ is set to $5$ and $20$ for comparison. We evaluate the prediction accuracy of the \emph{last} (i.e., next) item of each sequence in the testing set, similarly to~ \cite{hidasi2017recurrent}.

\subsection{Results Summary}
\label{resultsumm}
Overall performance results of all methods are illustrated in Table~\ref{overallresults5} and \ref{overallresults20}, which clearly show that the neural network models (i.e.,  \emph{Caser}, \emph{GRURec} and our model) obtain very promising accuracy in the top-$N$sequential recommendation task. For example, in \emph{MUSIC\_M5},  the three neural models perform more than $120$ times better on MRR@5 than \emph{MostPop}, which is a widely used recommendation benchmark. The best MRR@20 result we have achieved by \emph{NextItNet} is 0.3223 in this data set, which roughly means that
 the desired item is ranked on position 3  in average among the 20,000 candidate items.   
We then find that among these neural network based models, \emph{NextItNet} largely outperforms  \emph{Caser} $\&$ \emph{GRURec}. We believe there are several reasons contributing to the state-of-the-art results. First, as highlighted in Section \ref{conditionmodel}, \emph{NextItNet} takes full advantage of the complete sequential information. This can  be easily verified in Table \ref{withsubsession}, where we observe that \emph{Caser} $\&$ \emph{GRURec} without subsession perform extremely worse with long sessions. 
In addition, even with sub-session \emph{Caser} $\&$ \emph{GRURec} still show significantly worse results than \emph{NextItNet} because the separate optimization of each sub-session is clearly suboptimal compared with leveraging full sessions by  \emph{NextItNet}. 
Second, unlike \emph{Caser}, \emph{NextItNet} has no pooling layers, although it is also a CNN based model. As a result,  \emph{NextItNet} preserves the whole  spatial resolution of the original embedding matrix  $\bm{E}$ without any information lost. 
  The third advantage is that \emph{NextItNet} can support deeper layers by using residual learning, which better suits for modeling complicated relations and long-range dependencies. 
   We have separately validated the performance of residual block in Fig. \ref{residualreshape} (b) and showed the results in Table \ref{residual}.  It can be observed  that the performance of \emph{NextItNet} can be significantly improved by the residual block design. Table \ref{embedding} shows the impact of embedding sizes. 
\begin{figure*}
	\small
	\centering     
	\subfloat[\scriptsize  Loss ]{\label{yahoo-alphazero}\includegraphics[width=0.24\textwidth]{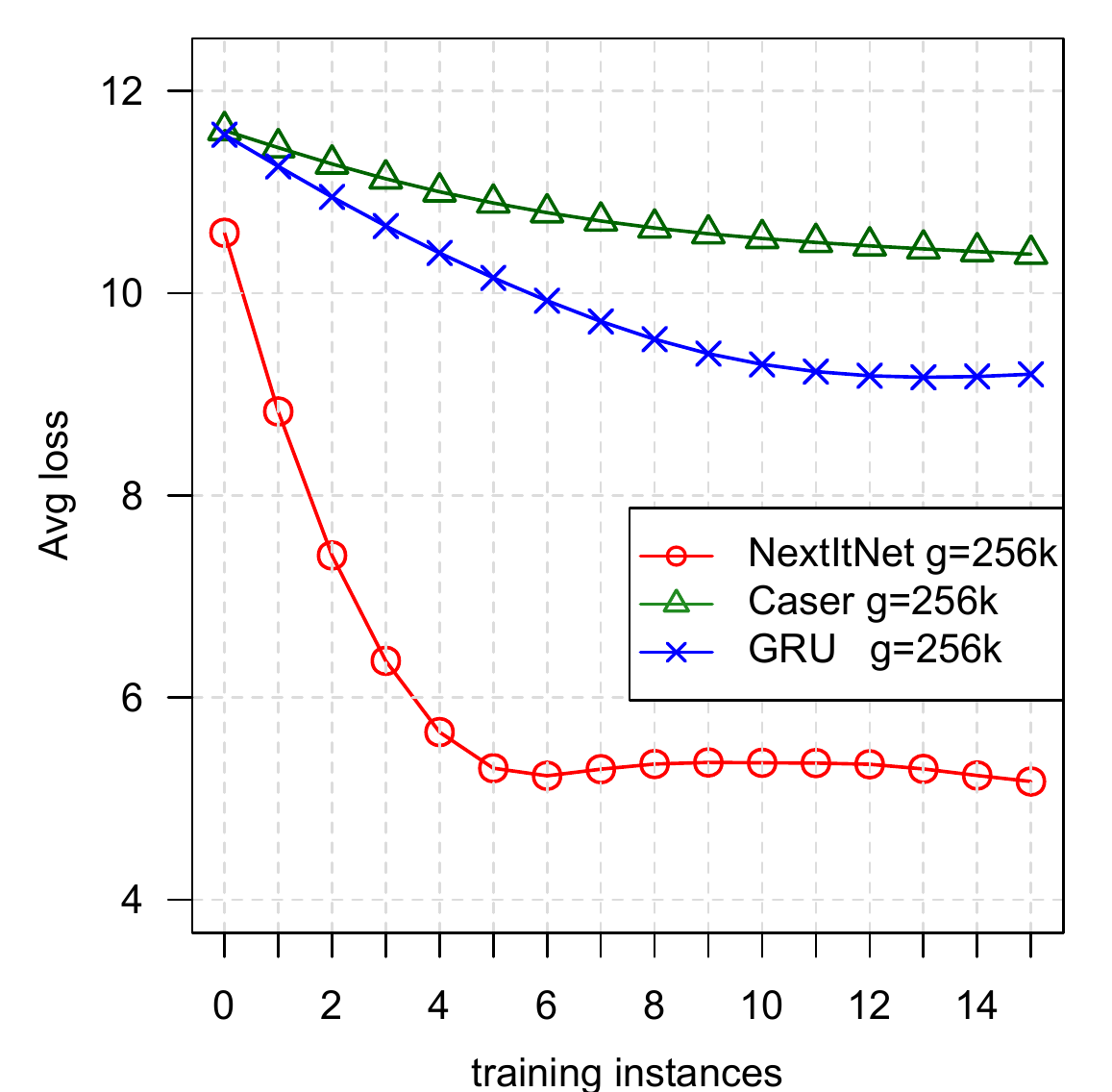}}
	\subfloat[\scriptsize  MRR@5]{\label{yahoo-alpha}\includegraphics[width=0.24\textwidth]{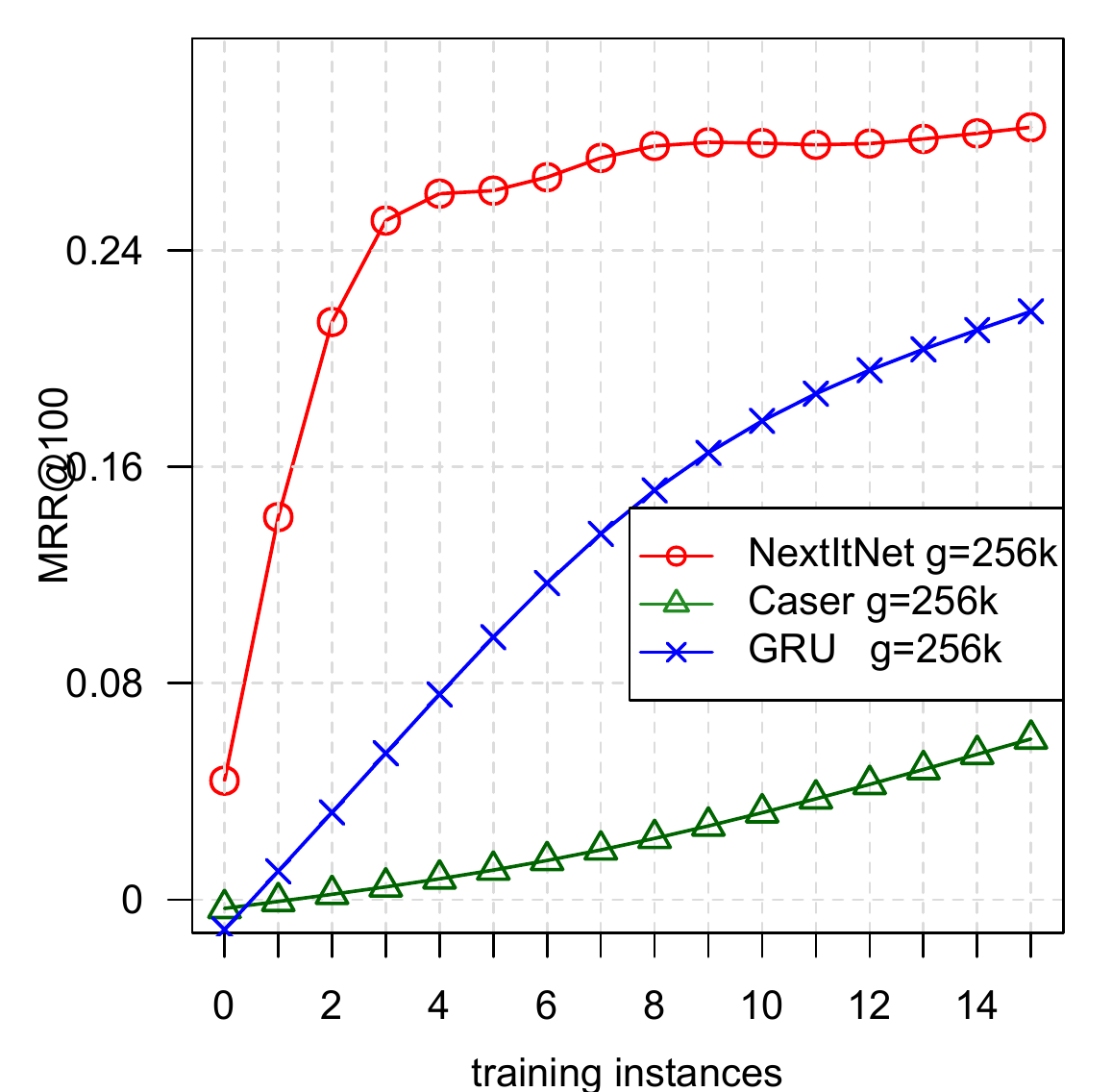}}
	\subfloat[\scriptsize  HR@5]{\label{yahoo-alpha}\includegraphics[width=0.24\textwidth]{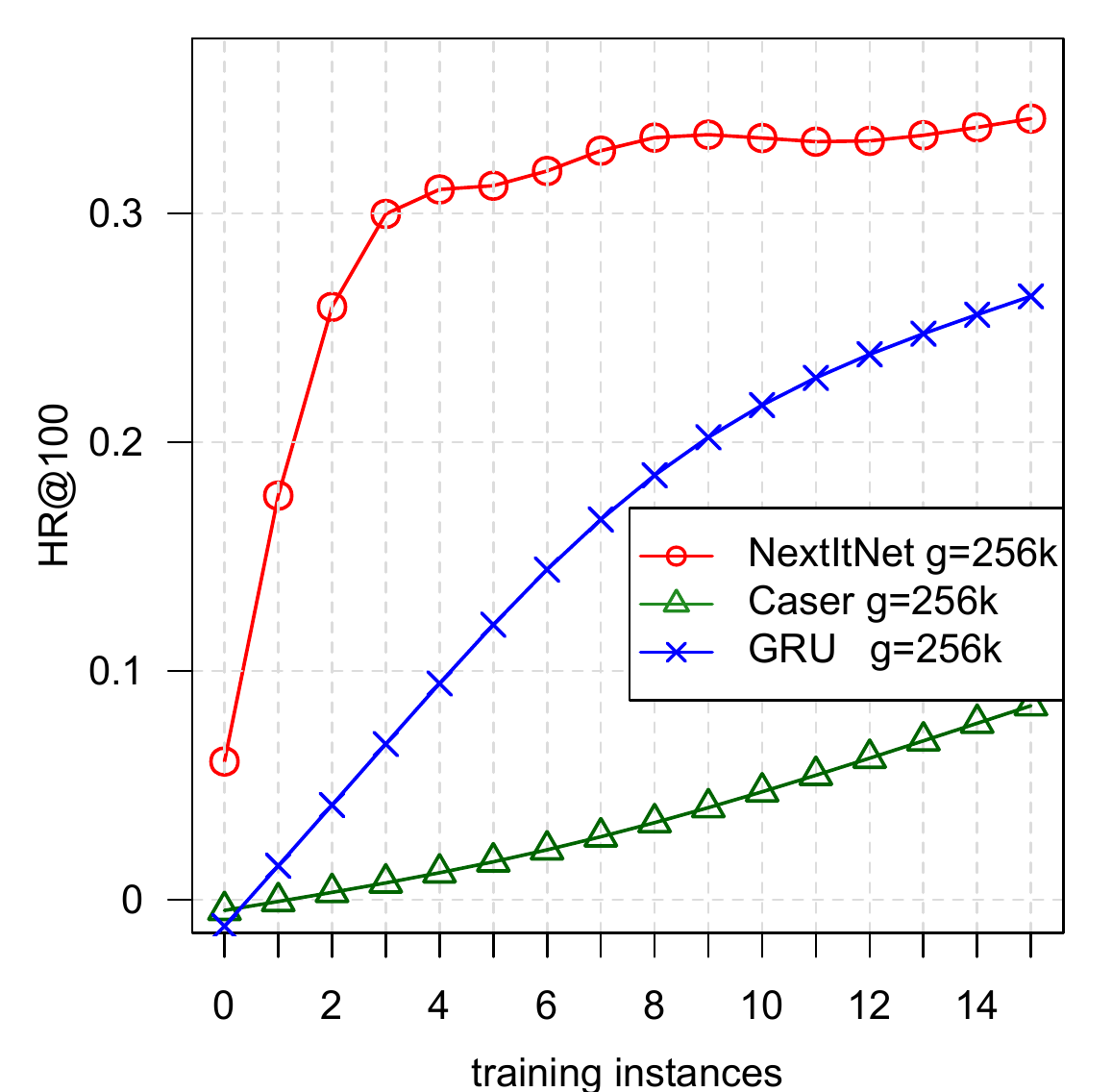}}
	\subfloat[\scriptsize  NDCG@5]{\label{yahoo-alpha}\includegraphics[width=0.24\textwidth]{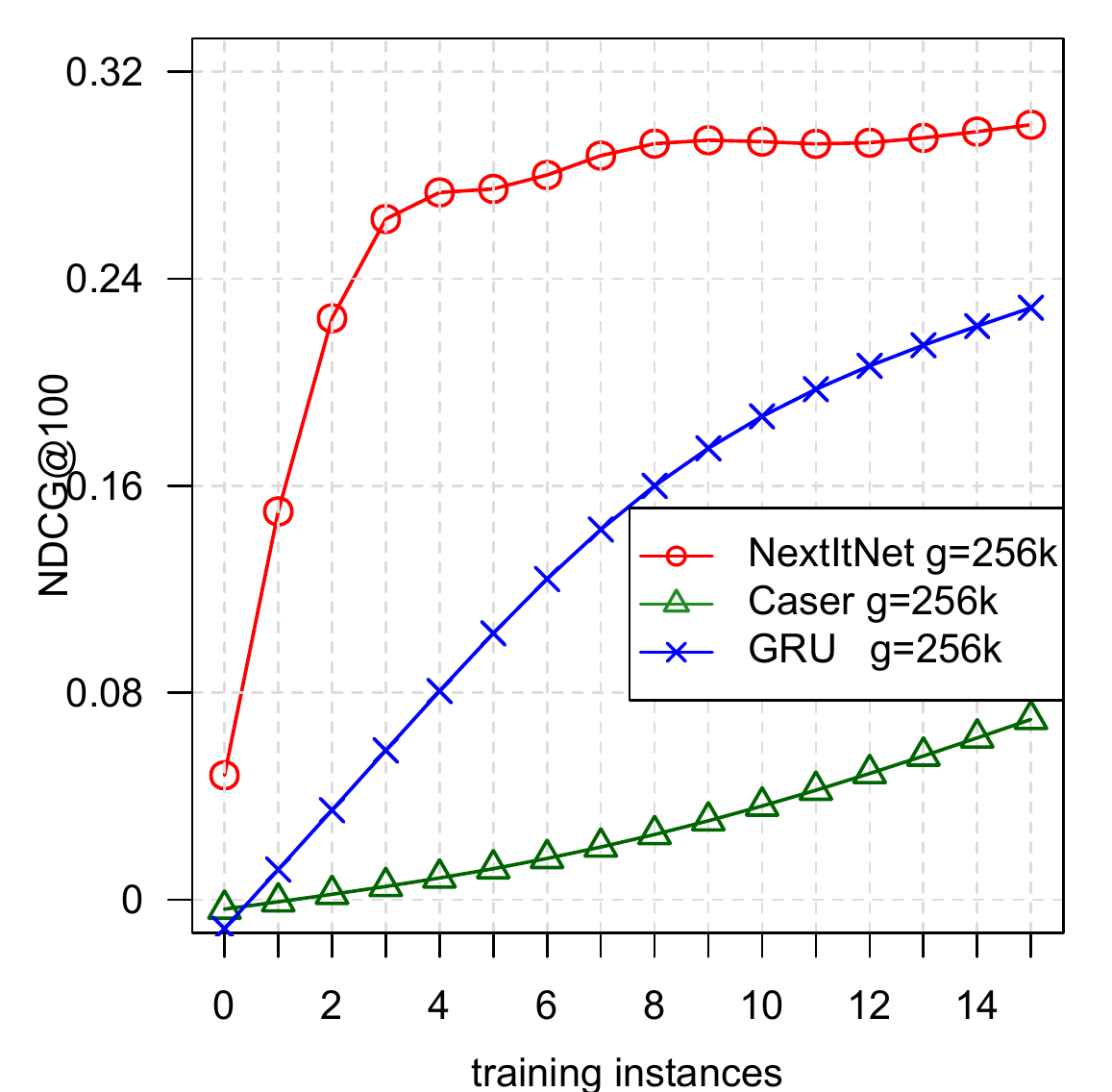}}
	\caption{\small Convergence behaviors of 	\emph{MUSIC\_L100}.  \emph{GRU} is short for \emph{GRURec}. $g=256$k means the number of training sequences (or sessions) of one unit in x-axis is 256k. Note that (1) to speed up the experiments, all of the convergence tests are evaluated on the first 1024 sessions in the testing set; (2) only \emph{NextItNet} has converged in above figures, while \emph{GRU} and \emph{Caser} require more training instances for convergence.} 
	\label{convergence100}
\end{figure*}

In addition to the advantage of recommendation accuracy, we have also evaluated the efficiency of \emph{NextItNet}  in Table~\ref{trainingtime}.
First, it can be seen that \emph{NextItNet} and \emph{Caser} requires less training time than \emph{GRURec}  in all three data sets. The reason that CNN-based models can be trained much faster is due to the full parallel mechanism of convolutions.
Clearly, the training speed advantage of CNN models are more preferred by modern parallel computing systems. Second, it shows that \emph{NextItNet} achieves further  improvements in training time compared with \emph{Caser}.
The faster training speed is mainly because \emph{NextItNet} leverages the complete sequential information during training and then converges much faster by less training epochs. To better understand of the convergence behaviours, we have shown them in Fig. \ref{convergence100}. As can be seen, our model with  the same number of training sessions converges faster (a) and better (b, c, d) than \emph {Caser} and \emph {GRURec}. This confirms our claim in Section \ref{conditionmodel} since \emph {Caser} and \emph {GRURec} cannot make full use of the internal sequential information in the session.


\begin{table} 
	\centering
	\caption{\small Overall training time  (mins).}
	\small
	\label{trainingtime}
	\begin{threeparttable}				
		\begin{tabular}{@{} L{18mm} C{18mm} C{18mm} C{18mm} @{}}
			\toprule
			\small Model   &  \small 	\emph{GRURec}  &  \small \emph{Caser}&  \small \emph{NextItNet}    \\
			\midrule
			\midrule
			\emph{MUSIC\_L5 }            &  66 &59 & 54 \\  
			\emph{MUSIC\_L20 }            &  282 &191 & 76  \\  
			\emph{MUSIC\_L100 }            &  586 &288 & 150  \\ 
			\bottomrule
		\end{tabular}
	\end{threeparttable}
\end{table}

\section{Conclusion}
In this paper, we presented a simple, efficient and highly effective convolutional generative model for session-based  top-$N$ item recommendations. 
The proposed model combines masked filters with 1D dilated convolutions to increase the receptive fields, which is very important to model the long-range dependencies. In addition, we have applied residual learning to enable training of much deeper networks. We have shown that our model can greatly outperform state-of-the-arts in real-world session-based recommendation tasks. The proposed model can serve as a generic method for modeling both short- and long-range session-based recommendation data.

For comparison purposes, we have not considered additional contexts in either our model or baselines. However, our  model is flexible to incorporate various context  information. For example, if we know the user identity $u$ and location $p$, the distribution in Eq.~(\ref{jointdis}) can be modified as follows to incorporate these information.
\begin{equation}
	\label{fusecontext}
		p(\textbf{x})=\prod_{i=1}^{t}p(x_i|x_{0:{i-1}},\textbf{u},\bm{P},\bm{\theta})p(x_0)
\end{equation}
where we can combine $\bm{E}$ (before convolution) or $\bm{E^o}$ (after convolution) with the user embedding vector $\textbf{u}$ and location matrix $\bm{P}$ by element-wise operations, such as multiplication, addition or concatenation.
 We  leave the evaluation for future work.

\begin{acks}
This paper was supported by the European Union's Horizon 2020 Research and Innovation program under grant agreement No 780787.

\end{acks}

\bibliographystyle{ACM-Reference-Format}

\bibliography{sample-bibliography} 


\begin{thebibliography}{34}


\ifx \showCODEN    \undefined \def \showCODEN     #1{\unskip}     \fi
\ifx \showDOI      \undefined \def \showDOI       #1{#1}\fi
\ifx \showISBNx    \undefined \def \showISBNx     #1{\unskip}     \fi
\ifx \showISBNxiii \undefined \def \showISBNxiii  #1{\unskip}     \fi
\ifx \showISSN     \undefined \def \showISSN      #1{\unskip}     \fi
\ifx \showLCCN     \undefined \def \showLCCN      #1{\unskip}     \fi
\ifx \shownote     \undefined \def \shownote      #1{#1}          \fi
\ifx \showarticletitle \undefined \def \showarticletitle #1{#1}   \fi
\ifx \showURL      \undefined \def \showURL       {\relax}        \fi
\providecommand\bibfield[2]{#2}
\providecommand\bibinfo[2]{#2}
\providecommand\natexlab[1]{#1}
\providecommand\showeprint[2][]{arXiv:#2}

\bibitem[\protect\citeauthoryear{Ba, Kiros, and Hinton}{Ba
  et~al\mbox{.}}{2016}]%
        {ba2016layer}
\bibfield{author}{\bibinfo{person}{Jimmy~Lei Ba}, \bibinfo{person}{Jamie~Ryan
  Kiros}, {and} \bibinfo{person}{Geoffrey~E Hinton}.}
  \bibinfo{year}{2016}\natexlab{}.
\newblock \showarticletitle{Layer normalization}.
\newblock \bibinfo{journal}{{\em arXiv preprint arXiv:1607.06450\/}}
  (\bibinfo{year}{2016}).
\newblock


\bibitem[\protect\citeauthoryear{Blanc and Rendle}{Blanc and Rendle}{2017}]%
        {blanc2017adaptive}
\bibfield{author}{\bibinfo{person}{Guy Blanc} {and} \bibinfo{person}{Steffen
  Rendle}.} \bibinfo{year}{2017}\natexlab{}.
\newblock \showarticletitle{Adaptive Sampled Softmax with Kernel Based
  Sampling}.
\newblock \bibinfo{journal}{{\em arXiv preprint arXiv:1712.00527\/}}
  (\bibinfo{year}{2017}).
\newblock


\bibitem[\protect\citeauthoryear{Chatzis, Christodoulou, and Andreou}{Chatzis
  et~al\mbox{.}}{2017}]%
        {Chatzis2017}
\bibfield{author}{\bibinfo{person}{Sotirios~P. Chatzis},
  \bibinfo{person}{Panayiotis Christodoulou}, {and} \bibinfo{person}{Andreas~S.
  Andreou}.} \bibinfo{year}{2017}\natexlab{}.
\newblock \showarticletitle{Recurrent Latent Variable Networks for
  Session-Based Recommendation}. In \bibinfo{booktitle}{{\em Proceedings of the
  2nd Workshop on Deep Learning for Recommender Systems}} {\em
  (\bibinfo{series}{DLRS 2017})}. \bibinfo{publisher}{ACM},
  \bibinfo{pages}{38--45}.
\newblock
\showISBNx{978-1-4503-5353-3}


\bibitem[\protect\citeauthoryear{Chen, Papandreou, Kokkinos, Murphy, and
  Yuille}{Chen et~al\mbox{.}}{2016}]%
        {chen2016deeplab}
\bibfield{author}{\bibinfo{person}{Liang-Chieh Chen}, \bibinfo{person}{George
  Papandreou}, \bibinfo{person}{Iasonas Kokkinos}, \bibinfo{person}{Kevin
  Murphy}, {and} \bibinfo{person}{Alan~L Yuille}.}
  \bibinfo{year}{2016}\natexlab{}.
\newblock \showarticletitle{Deeplab: Semantic image segmentation with deep
  convolutional nets, atrous convolution, and fully connected crfs}.
\newblock \bibinfo{journal}{{\em arXiv preprint arXiv:1606.00915\/}}
  (\bibinfo{year}{2016}).
\newblock


\bibitem[\protect\citeauthoryear{Cheng, Yang, Lyu, and King}{Cheng
  et~al\mbox{.}}{2013}]%
        {cheng2013you}
\bibfield{author}{\bibinfo{person}{Chen Cheng}, \bibinfo{person}{Haiqin Yang},
  \bibinfo{person}{Michael~R Lyu}, {and} \bibinfo{person}{Irwin King}.}
  \bibinfo{year}{2013}\natexlab{}.
\newblock \showarticletitle{Where You Like to Go Next: Successive
  Point-of-Interest Recommendation.}. In \bibinfo{booktitle}{{\em IJCAI}}.
\newblock


\bibitem[\protect\citeauthoryear{Cheng, Shen, Zhu, Kankanhalli, and Nie}{Cheng
  et~al\mbox{.}}{2017}]%
        {Zhiyong}
\bibfield{author}{\bibinfo{person}{Zhiyong Cheng}, \bibinfo{person}{Jialie
  Shen}, \bibinfo{person}{Lei Zhu}, \bibinfo{person}{Mohan Kankanhalli}, {and}
  \bibinfo{person}{Liqiang Nie}.} \bibinfo{year}{2017}\natexlab{}.
\newblock \showarticletitle{Exploiting Music Play Sequence for Music
  Recommendation}. In \bibinfo{booktitle}{{\em IJCAI}}.
\newblock


\bibitem[\protect\citeauthoryear{Cui, Wu, Huang, and Wang}{Cui
  et~al\mbox{.}}{2017}]%
        {CuiQiang}
\bibfield{author}{\bibinfo{person}{Qiang Cui}, \bibinfo{person}{Shu Wu},
  \bibinfo{person}{Yan Huang}, {and} \bibinfo{person}{Liang Wang}.}
  \bibinfo{year}{2017}\natexlab{}.
\newblock \showarticletitle{A Hierarchical Contextual Attention-based GRU
  Network for Sequential Recommendation}.
\newblock \bibinfo{journal}{{\em arXiv preprint arXiv:1711.05114\/}}
  (\bibinfo{year}{2017}).
\newblock


\bibitem[\protect\citeauthoryear{Gehring, Auli, Grangier, Yarats, and
  Dauphin}{Gehring et~al\mbox{.}}{2017}]%
        {gehring2017convolutional}
\bibfield{author}{\bibinfo{person}{Jonas Gehring}, \bibinfo{person}{Michael
  Auli}, \bibinfo{person}{David Grangier}, \bibinfo{person}{Denis Yarats},
  {and} \bibinfo{person}{Yann~N Dauphin}.} \bibinfo{year}{2017}\natexlab{}.
\newblock \showarticletitle{Convolutional Sequence to Sequence Learning}.
\newblock \bibinfo{journal}{{\em arXiv preprint arXiv:1705.03122\/}}
  (\bibinfo{year}{2017}).
\newblock


\bibitem[\protect\citeauthoryear{Gu, Lei, Barzilay, and Jaakkola}{Gu
  et~al\mbox{.}}{2016}]%
        {gu2016learning}
\bibfield{author}{\bibinfo{person}{Youyang Gu}, \bibinfo{person}{Tao Lei},
  \bibinfo{person}{Regina Barzilay}, {and} \bibinfo{person}{Tommi~S Jaakkola}.}
  \bibinfo{year}{2016}\natexlab{}.
\newblock \showarticletitle{Learning to refine text based recommendations.}. In
  \bibinfo{booktitle}{{\em EMNLP}}. \bibinfo{pages}{2103--2108}.
\newblock


\bibitem[\protect\citeauthoryear{He, Zhang, Ren, and Sun}{He
  et~al\mbox{.}}{2016a}]%
        {he2016deep}
\bibfield{author}{\bibinfo{person}{Kaiming He}, \bibinfo{person}{Xiangyu
  Zhang}, \bibinfo{person}{Shaoqing Ren}, {and} \bibinfo{person}{Jian Sun}.}
  \bibinfo{year}{2016}\natexlab{a}.
\newblock \showarticletitle{Deep residual learning for image recognition}. In
  \bibinfo{booktitle}{{\em CVPR}}. \bibinfo{pages}{770--778}.
\newblock


\bibitem[\protect\citeauthoryear{He, Zhang, Ren, and Sun}{He
  et~al\mbox{.}}{2016b}]%
        {he2016identity}
\bibfield{author}{\bibinfo{person}{Kaiming He}, \bibinfo{person}{Xiangyu
  Zhang}, \bibinfo{person}{Shaoqing Ren}, {and} \bibinfo{person}{Jian Sun}.}
  \bibinfo{year}{2016}\natexlab{b}.
\newblock \showarticletitle{Identity mappings in deep residual networks}. In
  \bibinfo{booktitle}{{\em ECCV}}. Springer, \bibinfo{pages}{630--645}.
\newblock


\bibitem[\protect\citeauthoryear{He and Chua}{He and Chua}{2017}]%
        {he2017neural}
\bibfield{author}{\bibinfo{person}{Xiangnan He} {and} \bibinfo{person}{Tat-Seng
  Chua}.} \bibinfo{year}{2017}\natexlab{}.
\newblock \showarticletitle{Neural factorization machines for sparse predictive
  analytics}. In \bibinfo{booktitle}{{\em SIGIR}}. ACM,
  \bibinfo{pages}{355--364}.
\newblock


\bibitem[\protect\citeauthoryear{He, He, Du, and Chua}{He
  et~al\mbox{.}}{2018}]%
        {he2018adversarial}
\bibfield{author}{\bibinfo{person}{Xiangnan He}, \bibinfo{person}{Zhankui He},
  \bibinfo{person}{Xiaoyu Du}, {and} \bibinfo{person}{Tat-Seng Chua}.}
  \bibinfo{year}{2018}\natexlab{}.
\newblock \showarticletitle{Adversarial personalized ranking for
  recommendation}. In \bibinfo{booktitle}{{\em SIGIR}}. ACM,
  \bibinfo{pages}{355--364}.
\newblock


\bibitem[\protect\citeauthoryear{Hidasi and Karatzoglou}{Hidasi and
  Karatzoglou}{2017}]%
        {hidasi2017recurrent}
\bibfield{author}{\bibinfo{person}{Bal{\'a}zs Hidasi} {and}
  \bibinfo{person}{Alexandros Karatzoglou}.} \bibinfo{year}{2017}\natexlab{}.
\newblock \showarticletitle{Recurrent Neural Networks with Top-k Gains for
  Session-based Recommendations}.
\newblock \bibinfo{journal}{{\em arXiv preprint arXiv:1706.03847\/}}
  (\bibinfo{year}{2017}).
\newblock


\bibitem[\protect\citeauthoryear{Hidasi, Karatzoglou, Baltrunas, and
  Tikk}{Hidasi et~al\mbox{.}}{2015}]%
        {hidasi2015session}
\bibfield{author}{\bibinfo{person}{Bal{\'a}zs Hidasi},
  \bibinfo{person}{Alexandros Karatzoglou}, \bibinfo{person}{Linas Baltrunas},
  {and} \bibinfo{person}{Domonkos Tikk}.} \bibinfo{year}{2015}\natexlab{}.
\newblock \showarticletitle{Session-based recommendations with recurrent neural
  networks}.
\newblock \bibinfo{journal}{{\em arXiv preprint arXiv:1511.06939\/}}
  (\bibinfo{year}{2015}).
\newblock


\bibitem[\protect\citeauthoryear{Ioffe and Szegedy}{Ioffe and Szegedy}{2015}]%
        {ioffe2015batch}
\bibfield{author}{\bibinfo{person}{Sergey Ioffe} {and}
  \bibinfo{person}{Christian Szegedy}.} \bibinfo{year}{2015}\natexlab{}.
\newblock \showarticletitle{Batch normalization: Accelerating deep network
  training by reducing internal covariate shift}. In \bibinfo{booktitle}{{\em
  ICML}}. \bibinfo{pages}{448--456}.
\newblock


\bibitem[\protect\citeauthoryear{Jean, Cho, Memisevic, and Bengio}{Jean
  et~al\mbox{.}}{2014}]%
        {jean2014using}
\bibfield{author}{\bibinfo{person}{S{\'e}bastien Jean},
  \bibinfo{person}{Kyunghyun Cho}, \bibinfo{person}{Roland Memisevic}, {and}
  \bibinfo{person}{Yoshua Bengio}.} \bibinfo{year}{2014}\natexlab{}.
\newblock \showarticletitle{On using very large target vocabulary for neural
  machine translation}.
\newblock \bibinfo{journal}{{\em arXiv preprint arXiv:1412.2007\/}}
  (\bibinfo{year}{2014}).
\newblock


\bibitem[\protect\citeauthoryear{Kalchbrenner, Espeholt, Simonyan, Oord,
  Graves, and Kavukcuoglu}{Kalchbrenner et~al\mbox{.}}{2016}]%
        {kalchbrenner2016neural}
\bibfield{author}{\bibinfo{person}{Nal Kalchbrenner}, \bibinfo{person}{Lasse
  Espeholt}, \bibinfo{person}{Karen Simonyan}, \bibinfo{person}{Aaron van~den
  Oord}, \bibinfo{person}{Alex Graves}, {and} \bibinfo{person}{Koray
  Kavukcuoglu}.} \bibinfo{year}{2016}\natexlab{}.
\newblock \showarticletitle{Neural machine translation in linear time}.
\newblock \bibinfo{journal}{{\em arXiv preprint arXiv:1610.10099\/}}
  (\bibinfo{year}{2016}).
\newblock


\bibitem[\protect\citeauthoryear{Larochelle and Murray}{Larochelle and
  Murray}{2011}]%
        {larochelle2011neural}
\bibfield{author}{\bibinfo{person}{Hugo Larochelle} {and} \bibinfo{person}{Iain
  Murray}.} \bibinfo{year}{2011}\natexlab{}.
\newblock \showarticletitle{The neural autoregressive distribution estimator}.
  In \bibinfo{booktitle}{{\em AISTATS}}. \bibinfo{pages}{29--37}.
\newblock


\bibitem[\protect\citeauthoryear{Li, Ren, Chen, Ren, Lian, and Ma}{Li
  et~al\mbox{.}}{2017}]%
        {li2017neural}
\bibfield{author}{\bibinfo{person}{Jing Li}, \bibinfo{person}{Pengjie Ren},
  \bibinfo{person}{Zhumin Chen}, \bibinfo{person}{Zhaochun Ren},
  \bibinfo{person}{Tao Lian}, {and} \bibinfo{person}{Jun Ma}.}
  \bibinfo{year}{2017}\natexlab{}.
\newblock \showarticletitle{Neural Attentive Session-based Recommendation}. In
  \bibinfo{booktitle}{{\em CIKM}}. ACM, \bibinfo{pages}{1419--1428}.
\newblock


\bibitem[\protect\citeauthoryear{Nair and Hinton}{Nair and Hinton}{2010}]%
        {nair2010rectified}
\bibfield{author}{\bibinfo{person}{Vinod Nair} {and}
  \bibinfo{person}{Geoffrey~E Hinton}.} \bibinfo{year}{2010}\natexlab{}.
\newblock \showarticletitle{Rectified linear units improve restricted boltzmann
  machines}. In \bibinfo{booktitle}{{\em ICML}}. \bibinfo{pages}{807--814}.
\newblock


\bibitem[\protect\citeauthoryear{Oord, Dieleman, Zen, Simonyan, Vinyals,
  Graves, Kalchbrenner, Senior, and Kavukcuoglu}{Oord et~al\mbox{.}}{2016a}]%
        {oord2016wavenet}
\bibfield{author}{\bibinfo{person}{Aaron van~den Oord}, \bibinfo{person}{Sander
  Dieleman}, \bibinfo{person}{Heiga Zen}, \bibinfo{person}{Karen Simonyan},
  \bibinfo{person}{Oriol Vinyals}, \bibinfo{person}{Alex Graves},
  \bibinfo{person}{Nal Kalchbrenner}, \bibinfo{person}{Andrew Senior}, {and}
  \bibinfo{person}{Koray Kavukcuoglu}.} \bibinfo{year}{2016}\natexlab{a}.
\newblock \showarticletitle{Wavenet: A generative model for raw audio}.
\newblock \bibinfo{journal}{{\em arXiv preprint arXiv:1609.03499\/}}
  (\bibinfo{year}{2016}).
\newblock


\bibitem[\protect\citeauthoryear{Oord, Kalchbrenner, and Kavukcuoglu}{Oord
  et~al\mbox{.}}{2016b}]%
        {oord2016pixel}
\bibfield{author}{\bibinfo{person}{Aaron van~den Oord}, \bibinfo{person}{Nal
  Kalchbrenner}, {and} \bibinfo{person}{Koray Kavukcuoglu}.}
  \bibinfo{year}{2016}\natexlab{b}.
\newblock \showarticletitle{Pixel recurrent neural networks}.
\newblock \bibinfo{journal}{{\em arXiv preprint arXiv:1601.06759\/}}
  (\bibinfo{year}{2016}).
\newblock


\bibitem[\protect\citeauthoryear{Oord, Kalchbrenner, Vinyals, Espeholt, Graves,
  and Kavukcuoglu}{Oord et~al\mbox{.}}{2016c}]%
        {oord2016conditional}
\bibfield{author}{\bibinfo{person}{A{\"a}ron van~den Oord},
  \bibinfo{person}{Nal Kalchbrenner}, \bibinfo{person}{Oriol Vinyals},
  \bibinfo{person}{Lasse Espeholt}, \bibinfo{person}{Alex Graves}, {and}
  \bibinfo{person}{Koray Kavukcuoglu}.} \bibinfo{year}{2016}\natexlab{c}.
\newblock \showarticletitle{Conditional image generation with pixelcnn
  decoders}. In \bibinfo{booktitle}{{\em NIPS}}. Curran Associates Inc.,
  \bibinfo{pages}{4797--4805}.
\newblock


\bibitem[\protect\citeauthoryear{Quadrana, Karatzoglou, Hidasi, and
  Cremonesi}{Quadrana et~al\mbox{.}}{2017}]%
        {quadrana2017personalizing}
\bibfield{author}{\bibinfo{person}{Massimo Quadrana},
  \bibinfo{person}{Alexandros Karatzoglou}, \bibinfo{person}{Bal{\'a}zs
  Hidasi}, {and} \bibinfo{person}{Paolo Cremonesi}.}
  \bibinfo{year}{2017}\natexlab{}.
\newblock \showarticletitle{Personalizing Session-based Recommendations with
  Hierarchical Recurrent Neural Networks}.
\newblock \bibinfo{journal}{{\em arXiv preprint arXiv:1706.04148\/}}
  (\bibinfo{year}{2017}).
\newblock


\bibitem[\protect\citeauthoryear{Sercu and Goel}{Sercu and Goel}{2016}]%
        {sercu2016dense}
\bibfield{author}{\bibinfo{person}{Tom Sercu} {and} \bibinfo{person}{Vaibhava
  Goel}.} \bibinfo{year}{2016}\natexlab{}.
\newblock \showarticletitle{Dense Prediction on Sequences with Time-Dilated
  Convolutions for Speech Recognition}.
\newblock \bibinfo{journal}{{\em arXiv preprint arXiv:1611.09288\/}}
  (\bibinfo{year}{2016}).
\newblock


\bibitem[\protect\citeauthoryear{Smirnova and Vasile}{Smirnova and
  Vasile}{2017}]%
        {Elena}
\bibfield{author}{\bibinfo{person}{Elena Smirnova} {and}
  \bibinfo{person}{Flavian Vasile}.} \bibinfo{year}{2017}\natexlab{}.
\newblock \showarticletitle{Contextual Sequence Modeling for Recommendation
  with Recurrent Neural Networks}.
\newblock \bibinfo{journal}{{\em arXiv preprint arXiv:1706.07684\/}}
  (\bibinfo{year}{2017}).
\newblock


\bibitem[\protect\citeauthoryear{Tan, Xu, and Liu}{Tan et~al\mbox{.}}{2016}]%
        {tan2016improved}
\bibfield{author}{\bibinfo{person}{Yong~Kiam Tan}, \bibinfo{person}{Xinxing
  Xu}, {and} \bibinfo{person}{Yong Liu}.} \bibinfo{year}{2016}\natexlab{}.
\newblock \showarticletitle{Improved recurrent neural networks for
  session-based recommendations}. In \bibinfo{booktitle}{{\em Proceedings of
  the 1st Workshop on Deep Learning for Recommender Systems}}. ACM,
  \bibinfo{pages}{17--22}.
\newblock


\bibitem[\protect\citeauthoryear{Tang and Wang}{Tang and Wang}{2018}]%
        {tang2018caser}
\bibfield{author}{\bibinfo{person}{Jiaxi Tang} {and} \bibinfo{person}{Ke
  Wang}.} \bibinfo{year}{2018}\natexlab{}.
\newblock \showarticletitle{Personalized Top-N Sequential Recommendation via
  Convolutional Sequence Embedding}. In \bibinfo{booktitle}{{\em ACM
  International Conference on Web Search and Data Mining}}.
\newblock


\bibitem[\protect\citeauthoryear{Tuan and Phuong}{Tuan and Phuong}{2017}]%
        {Tuan:2017:CNS:3109859.3109900}
\bibfield{author}{\bibinfo{person}{Trinh~Xuan Tuan} {and}
  \bibinfo{person}{Tu~Minh Phuong}.} \bibinfo{year}{2017}\natexlab{}.
\newblock \showarticletitle{3D Convolutional Networks for Session-based
  Recommendation with Content Features}. In \bibinfo{booktitle}{{\em RecSys}}.
  \bibinfo{publisher}{ACM}.
\newblock


\bibitem[\protect\citeauthoryear{Yu and Koltun}{Yu and Koltun}{2015}]%
        {yu2015multi}
\bibfield{author}{\bibinfo{person}{Fisher Yu} {and} \bibinfo{person}{Vladlen
  Koltun}.} \bibinfo{year}{2015}\natexlab{}.
\newblock \showarticletitle{Multi-scale context aggregation by dilated
  convolutions}.
\newblock \bibinfo{journal}{{\em arXiv preprint arXiv:1511.07122\/}}
  (\bibinfo{year}{2015}).
\newblock


\bibitem[\protect\citeauthoryear{Yuan, Guo, Jose, Chen, Yu, and Zhang}{Yuan
  et~al\mbox{.}}{2016}]%
        {yuan2016lambdafm}
\bibfield{author}{\bibinfo{person}{Fajie Yuan}, \bibinfo{person}{Guibing Guo},
  \bibinfo{person}{Joemon~M Jose}, \bibinfo{person}{Long Chen},
  \bibinfo{person}{Haitao Yu}, {and} \bibinfo{person}{Weinan Zhang}.}
  \bibinfo{year}{2016}\natexlab{}.
\newblock \showarticletitle{Lambdafm: learning optimal ranking with
  factorization machines using lambda surrogates}. In \bibinfo{booktitle}{{\em
  CIKM}}. ACM, \bibinfo{pages}{227--236}.
\newblock


\bibitem[\protect\citeauthoryear{Yuan, Guo, Jose, Chen, Yu, and Zhang}{Yuan
  et~al\mbox{.}}{2017}]%
        {yuan2017boostfm}
\bibfield{author}{\bibinfo{person}{Fajie Yuan}, \bibinfo{person}{Guibing Guo},
  \bibinfo{person}{Joemon~M Jose}, \bibinfo{person}{Long Chen},
  \bibinfo{person}{Haitao Yu}, {and} \bibinfo{person}{Weinan Zhang}.}
  \bibinfo{year}{2017}\natexlab{}.
\newblock \showarticletitle{Boostfm: Boosted factorization machines for top-n
  feature-based recommendation}. In \bibinfo{booktitle}{{\em IUI}}. ACM,
  \bibinfo{pages}{45--54}.
\newblock


\bibitem[\protect\citeauthoryear{Yuan, Xin, He, Guo, Zhang, Chua, and
  Jose}{Yuan et~al\mbox{.}}{2018}]%
        {yuan2018fbgd}
\bibfield{author}{\bibinfo{person}{Fajie Yuan}, \bibinfo{person}{Xin Xin},
  \bibinfo{person}{Xiangnan He}, \bibinfo{person}{Guibing Guo},
  \bibinfo{person}{Weinan Zhang}, \bibinfo{person}{Tat-Seng Chua}, {and}
  \bibinfo{person}{Joemon Jose}.} \bibinfo{year}{2018}\natexlab{}.
\newblock \showarticletitle{fBGD: Learning Embeddings From Positive Unlabeled
  Data with BGD}.
\newblock \bibinfo{journal}{{\em UAI\/}}.
\newblock


\end{thebibliography}
\scriptsize
\end{document}